\documentclass[lettersize,journal]{IEEEtran}
\usepackage{amsmath,amsfonts}
\usepackage{array}
\usepackage[caption=false,font=normalsize,labelfont=sf,textfont=sf]{subfig}
\usepackage{textcomp}
\usepackage{stfloats}
\usepackage{amsmath}
\usepackage{eqnarray,amsmath}
\usepackage{url}
\usepackage{amssymb}

\usepackage{amssymb}
\usepackage{lipsum}
\usepackage{cite}
\usepackage[center]{caption}
\usepackage{verbatim}
\usepackage{graphicx}
\usepackage[table]{xcolor}
\usepackage{amsfonts}
\usepackage{hyperref}
\usepackage[font=footnotesize]{caption}
\usepackage[left=0.39in,right=0.39in,top=0.48in,bottom=0.78in]{geometry}

\def\BibTeX{{\rm B\kern-.05em{\sc i\kern-.025em b}\kern-.08em
    T\kern-.1667em\lower.7ex\hbox{E}\kern-.125emX}}
\usepackage{balance}
\usepackage{algorithm}
\usepackage{algpseudocode}
\begin{document}
\title{Beamformed Fingerprint-Based Transformer Network for Trajectory Estimation and Path Determination in Outdoor mmWave MIMO Systems}
\author{Mohammad Shamsesalehi, Mahmoud Ahmadian Attari, Mohammad Amin Maleki Sadr, Benoit Champagne

\thanks{Copyright (c) 20xx IEEE. Personal use of this material is permitted. However, permission to use this material for any other purposes must be obtained from the IEEE by sending a request to pubs-permissions@ieee.org.

Mohammad Shamsesalehi and Mahmoud Ahmadian Attari are with the Faculty of Electrical Engineering, K. N. Toosi University of Technology, Tehran, Iran. (e-mail: shamsesalehi@email.kntu.ac.ir, mahmoud@eetd.kntu.ac.ir),
Mohammad Amin Maleki Sadr is a senior research scientist at 
Huawei, Toronto, Canada. (e-mail: mohammadamin.malekisadr@huawei.com),
Benoit Champagne is with the Department of Electrical and Computer Engineering, McGill University, Montréal, Canada (e-mail: benoit.champagne@mcgill.ca).}}
\maketitle

\begin{abstract}
Radio transmissions in millimeter wave (mmWave) bands have gained significant interest for applications demanding precise device localization and trajectory estimation. 
This paper explores novel neural network (NN) architectures suitable for trajectory estimation and path determination in a mmWave multiple-input multiple-output (MIMO) outdoor system based on localization data from beamformed fingerprint (BFF). The NN architecture captures sequences of BFF signals from different users, and through the application of learning mechanisms, subsequently estimate their trajectories.
In turn, this information is employed to find the shortest path to the target, thereby enabling more efficient navigation.  Specifically, we propose a two-stage procedure for trajectory estimation and optimal path finding. In the first stage, a transformer network (TN) based on attention mechanisms is developed to predict trajectories of wireless devices using BFF sequences captured in a mmWave MIMO outdoor system. In the second stage, a novel algorithm based on Informed Rapidly-exploring Random Trees (iRRT*) is employed to determine the optimal path to target locations using trajectory estimates derived in the first stage. The effectiveness of the proposed schemes is validated through numerical experiments, using a comprehensive dataset of radio measurements, generated using ray tracing simulations to model outdoor propagation at 28 GHz. We show that our proposed TN-based trajectory estimator outperforms other methods from the recent literature and can successfully generalize to new trajectories outside the training set. Furthermore, our proposed iRRT* algorithm is able to consistently provide the shortest path to the target.

\end{abstract}


\section{Introduction}
\IEEEPARstart{B}{eyond} 5G networks operating in millimeter wave (mmWave) bands will provide unprecedented bandwidth, while achieving unmatched performance in terms of throughput, latency, and reliability over a wide range of transmission distances \cite{7413967}. 
Reaching these promised milestones will require both the transmitter and receiver equipment to have precise information about the relative position and orientation of each other \cite{mogyorosi2022positioning}. Furthermore, accurate localization information is essential for several key applications, including trajectory estimation and path determination \cite{yin2022millimeter}.

From the perspective of radio environment, localization in mmWave bands can be broadly classified in terms of indoor and outdoor scenarios. A representative subset of works on indoor localization includes \cite{vukmirovic2018position, lin20183, vashist2021kf}. 
In \cite{vukmirovic2018position},  the authors introduce a massive MIMO architecture with distributed steerable subarrays for line-of-sight localization in mmWave, along with a multi-stage search strategy to resolve problems posed by high side-lobe levels.
In \cite{lin20183}, a hybrid positioning scheme based on signal strength and angle of arrival (AoA) is proposed for a mmWave m-MIMO system with a large-scale cylindrical array, wherein channel compression and beamspace transformation are employed to reduce AoA estimation complexity and enable accurate 3D target positioning.
In \cite{vashist2021kf}, the authors propose an integrated Kalman filter (KF) and machine learning (ML) localization system, wherein KF and ML synergistically contribute to improving indoor positioning accuracy. 
Representative works on outdoor localization, which is the focus of this work, include \cite{shen20212d, sadr2021uncertainty, mendrzik2019localization}. In \cite{shen20212d}, the authors propose a fingerprint-based cooperative approach leveraging information theoretic concepts for robust localization in mmWave massive MIMO systems. In \cite{sadr2021uncertainty}, the Monte Carlo (MC) dropout-based method is proposed for capturing the uncertainty in a CNN-based mmWave MIMO outdoor localization system. 
In \cite{mendrzik2019localization}, the statistical dependency between channel state information (CSI) and the states of a user device (i.e., position, orientation, and clock offset) is leveraged to provide situational awareness by inferring the device's state and a map of its propagation environment.

From a design perspective, localization methods can be classified in terms of their modeling approach, i.e., physical versus data-driven. Methods in the first class rely on physical models to recast the behavior of radio signal propagation and are dependent on knowledge of the environment and channel characteristics \cite{koivisto2017joint, koivisto2017high}. However, since channel models at mmWave frequencies rely on complex phenomena and numerous parameters that cannot be easily determined, this approach poses significant challenges for application in urban radio environments \cite{kanhere2018position}. In contrast, data-driven models for localization in mmWave MIMO systems estimate user position by collecting radio data or \emph{fingerprints} for a large set of representative positions, and then employing ML algorithms to predict the location of a new target based on its radio imprints \cite{8307353, 10292876}.
In \cite{fan2023fast}, the authors introduce a fast direct localization solver to address the computational challenges of the common two-step triangulation-based approach. This solver utilizes the deep Alternating Direction Method of Multipliers (ADMM), which significantly improves accuracy and reduces complexity.
In  \cite{gante2018beamformed}, a beamformed fingerprint (BFF) method is proposed for precise outdoor positioning by leveraging mmWave MIMO transmissions to achieve steerable and highly focused radiation patterns. In effect, BFF employs a two-dimensional matrix containing sequential information along both spatial and temporal dimensions.

The ML approaches can effectively model complex nonlinearities and achieve high prediction accuracy when fed by rich information data. Hence, selecting the appropriate measurable features is one of the main challenges in fingerprint schemes \cite{ye2017neural}. In \cite{zhang2017path}, an enhanced fingerprint-based localization method that incorporates a path loss model for fingerprint generation and localization is introduced. 
In \cite{zhou2017robust}, another localization method is developed, which employs crowd-sourcing to gather probabilistic radio signal strength (RSS) data from numerous smartphones and create a database of fingerprints.
A related CSI-based approach is introduced in \cite{pecoraro2018csi}, where signal fingerprints describe the shape of the channel frequency response.

Besides the use of channel measurements to determine the position of user equipments, extensive research has been conducted on their trajectory estimation based on physical approaches. In \cite{9500482}, a map-based positioning algorithm is tested in an outdoor urban microcell environment using data collected at 142 GHz. In this study, an extended Kalman filter tracks the user equipment’s position and velocity along a rectangular path using AoA and time of flight measurements. In \cite{7848938}, the unscented Kalman filter and extended Kalman filter are employed to jointly determine the position of a mobile user and perform network synchronization. In \cite{8440914}, the authors employ an extended Kalman filter to track the 3D position of user equipments, by exploiting information derived from beamformed reference signals transmitted by the base station at regular intervals.

Using sequences of BFFs within a data-driven framework represents an especially attractive means to estimate user trajectories in urban environments, which can be achieved by employing deep learning
(DL) architectures.
While the recurrent neural network (RNN) is frequently utilized to model sequential data, its gradient computation is plagued with various issues (explosion, vanishing mode, etc.), so that the model's effectiveness, in particular its ability to remember history, diminishes over time \cite{lalapura2024systematic}. These issues can be addressed in part by incorporating gating mechanisms in the RNN cells, with long short-term memory (LSTM) and gated recurrent units (GRUs) being the most common architectures \cite{7508408}. In \cite{bai2018rfedrnn}, such an RNN-based network is used to learn mobility patterns from radio frequency (RF) fingerprints in an indoor scenario. Recently, transformer networks (TNs), also known as \emph{transformers}, have demonstrated superior performance and flexibility in capturing and modeling data with a sequential nature \cite{vaswani2017attention}. Nevertheless, to the best of our knowledge, their potential application to BFF-based trajectory estimation remains an open problem.

Path determination in outdoor environments is a crucial extension of trajectory estimation, involving the synthesis of a collision-free path from an initial to a target position with minimum cost. Different criteria, such as smoothness and low energy consumption, can influence the determination of an optimal path \cite{noreen2016optimal}. Two main categories of path determination algorithms based on discrete search can be identified, namely: \emph{grid-based} and \emph{sampling-based}. Grid-based algorithms, like D*, wavefront, Dijkstra, Phi*, and A * \cite{stentz1994optimal, zelinsky1993planning, dijkstra1959, nash2009, zhan2014efficient, quan2020survey}, guarantee completeness and optimality but face challenges in large areas due to exponential growth in search space and time. Sampling-based algorithms, like probabilistic road map (PRM) and rapidly exploring random trees (RRT) \cite{he2021asynchronous, choudhary2023sampling}, partly address these challenges, with PRM lacking optimality and RRT focusing on target orientation. Several variants of RRT have been proposed, including RRT* which introduces pruning optimization for asymptotic optimality \cite{karaman2011sampling}, RRT*-smart, and RRT\# which aim to enhance convergence \cite{nasir2013rrt, arslan2013use}, and Informed RRT* (iRRT*), which constrains sampling intervals for further improved convergence \cite{gammell2014informed}. Additionally, reinforcement learning and evolutionary algorithms (e.g., genetic algorithm, ant colony optimization, particle swarm optimization, artificial bee colony) offer alternative approaches for multi-objective problems. However, optimality is not guaranteed, and computational complexity remains high \cite{10125088, arora2014robotic, dorigo2006ant, sameshima2014strrt, liang2015efficient}.

In this paper, we introduce a BFF-based DL network for trajectory estimation and path determination in a mmWave MIMO outdoor system. We leverage the BFF and a TN-based architecture to capture long-term data for trajectory estimation. Our approach is supplemented with an efficient path determination scheme. While satellite-based technologies can provide outdoor location services, they may underperform in information-deprived situations. Our BFF-based method in the mmWave band offers precise, rich information for more accurate estimation and path determination. The main innovative contributions of this work are summarized below:\footnote{An earlier version of this work limited to trajectory estimation for short BFF sequences was presented in \cite{shamsesalehi2024bff}.}
\begin{itemize}
\item  We propose a two-stage DL-based architecture for
trajectory estimation and optimal path determination based on localization data. Specifically, the proposed framework allows the capture of long trajectories based on rich information derived from BFF sequences in mmWave MIMO outdoor systems.

\item In the first stage, a TN-based attention mechanism is developed to predict trajectories of wireless user devices using the captured BFF sequences. The proposed TN architecture enables the identification of small changes in the direction of users, while for long sequences, parallel processing allows faster and more efficient computation of user trajectories.

\item In the second stage, a key novelty of our approach lies in adapting the iRRT* algorithm to incorporate trajectory estimates derived from the first stage. This integration enables dynamic and predictive path determination, overcoming the limitations of traditional methods that rely on static or pre-determined inputs.

\item By leveraging trajectory estimation, the proposed method reduces computational overhead while improving the adaptability and precision of the path determination process, ensuring higher quality outcomes.

\end{itemize}

The validity of the proposed schemes is demonstrated through numerical experiments, using a large dataset of radio measurements in an outdoor environment, supplemented with ray tracing to simulate wireless propagation at 28 GHz.  We show that our proposed TN-based trajectory estimator outperforms other methods from the recent literature and can successfully generalize to new trajectories outside the training set. Furthermore, our proposed iRRT* algorithm can consistently provide the shortest path to the target position. Table \ref{t0} provides a comparison of related works with the proposed approach, highlighting their methodologies, capabilities in localization, trajectory estimation (TE), and path determination (PD), as well as the key contribution of addressed by each study.

\begin{table*}[ht]
\centering
\caption{Comparison of Related Works with the Proposed Approach}
\label{tab:comparison}
\begin{tabular}{|c|l|c|c|c|p{6cm}|}
\hline
\textbf{Ref.} & \textbf{Approach} & \textbf{Localization} & \textbf{TE} & \textbf{PD} & \textbf{Key Contribution} \\
\hline
\cite{vukmirovic2018position} & Physical indoor model localization & Yes & No & No & Resolves line-of-sight localization challenges in mmWave environments. \\
\hline
\cite{bai2018rfedrnn} & RF fingerprint-based user tracking & No & Yes & No & Addresses learning mobility patterns in indoor scenarios. \\
\hline
\cite{gante2018beamformed} & Beamforming fingerprint (BFF)-based localization & Yes & No & No & Focuses on deep localization techniques. \\
\hline
\cite{gante2020deep} & BFF-based localization and trajectory estimation & Yes & Yes & Yes & Combines deep localization and trajectory estimation. \\
\hline
\cite{9626568} & Localization in mmWave band & Yes & No & No & Provides uncertainty bounds for localization. \\
\hline
\cite{gammell2014informed} & Informed sampling-based algorithm & No & No & Yes & Enhances efficient path determination. \\
\hline
\cite{shamsesalehi2024bff} & BFF-based trajectory estimation & Yes & Yes & No & Improves deep trajectory estimation efficiency. \\
\hline
\textbf{This Work} & BFF-based approach combined with sampling-based methods & Yes & Yes & Yes & Achieves efficient deep trajectory estimation and path determination. \\
\hline
\end{tabular}
\label{t0}
\end{table*}

The rest of the paper is organized as follows: In Section II, the system model for BFF-based localization in a mmWave MIMO outdoor system is exposed. In Section III, the attention-based mechanism (i.e., TN) for the BFF-based trajectory estimation problem is introduced. In Section IV, the novel iRRT*-based path determination approach is developed. In Section V, simulation results are presented. Finally, Section VI concludes the paper.

\section{System Model and Data Acquisition}

As outdoor 5G base stations (BSs) are typically positioned at elevated locations within urban environments, buildings emerge as the primary obstacles that remain stationary over extended periods. Under such conditions, the received power delay profile (PDP) measurements, as detailed in \cite{guan2020channel}, are anticipated to provide significant benefit for radio localization. To maximize coverage, BSs can employ a sequence of directive beamforming (BF) patterns to transmit signals, effectively covering all feasible transmission angles. Consequently, receivers can capture multiple distinct PDPs corresponding to the different beampatterns to enrich the data set. Because of the intricate propagation mechanisms that occur when obstacles are present, it is expected that the measured PDPs will exhibit significant variations or discontinuities across the localization area. These abrupt variations offer valuable
spatial information to a trained localization model. Moreover, the PDP data is critical for trajectory estimation and path determination, providing detailed spatial-temporal insights into the mmWave environment, including multipath delays and power levels. This data enhances the model accuracy by capturing key environmental factors like obstacles. Additionally, PDP data reduces input complexity compared to raw channel impulse responses, improving computational efficiency while maintaining result quality \cite{gante2020deep}.

\begin{figure}
\centering
\includegraphics[width=0.4\textwidth]{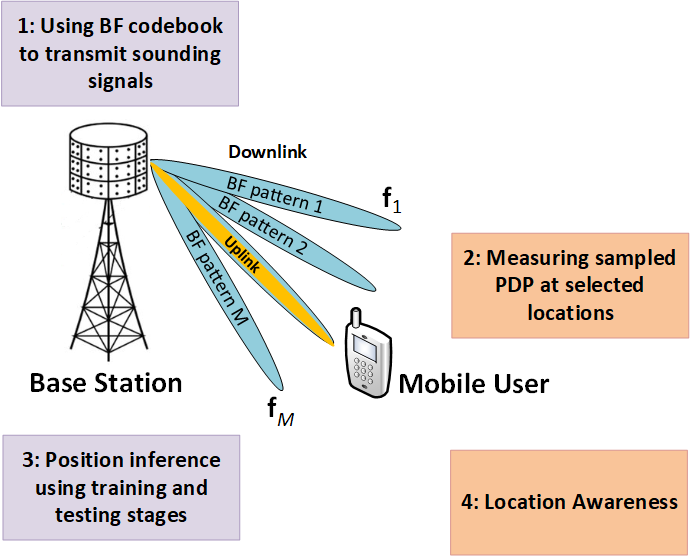}
\caption{\label{fig:frog} System model of BFF-based  MIMO mmWave localization }
\label{system_model}
\end{figure}

The process of obtaining the BFF data through the use of a BF codebook at the BS and PDP monitoring at the mobile user is illustrated in Fig. \ref{system_model}. At the BS, directional sounding signals are generated to facilitate the extraction of BFF data from the radio environment for both purposes of offline model training and online prediction, i.e., trajectory estimation. 
During the training phase, the BS equipped with $N_T$ antennas transmits signals within a specific frequency band $B$ using a predetermined set of $M$ transmit BF vectors from a codebook $\mathcal{C} = \left\{ \textbf{f}_1, \textbf{f}_2, \cdots , \textbf{f}_M \right\}$, where $\textbf{f}_i\in \mathbb{C}^{N_T\times 1}$ denotes the $i$th BF vector. 
A user terminal equipped with $N_R$ antennas and located at a known position (to be varied) is used to capture the sequence of transmitted signals. The received user signal from the $i$th codebook at a specific sounding frequency $f \in B$, can be expressed as:
 \begin{equation}   \label{eq1}
r_i(f)= \textbf{w}^{H} \left( \textbf{H}\;\textbf{f}_i\;s+\textbf{z}_{i} \right),\;\;\; i\in \left\{ 1,\cdots ,M \right\},
\end{equation}
where $s\in \mathbb{C}$ is the sounding signal amplitude, $\textbf{H}\in \mathbb{C}^{N_R\times N_T}$ is the complex channel gain matrix between the BS and user device, $\textbf{w}\in \mathbb{C}^{N_R\times 1}$ is the BF vector at the receiver and $\textbf{z}_{i}\in \mathbb{C}^{N_R\times 1}$ is an additive noise term. 

At the user terminal, through processing of the received signals $r_i(f)$ over all frequencies $f \in B$, the received power is measured and an overall PDP $P_i(\tau)$ is calculated, where $\tau$ represents the (continuous) delay variable. A discrete-time version of the PDP, denoted as $P_i[j]$, is obtained by uniformly sampling $P_i(\tau)$ at frequency $F_s = {T_s}^{-1}$, where $T_s$ denotes the sampling interval \cite{guan2020channel}. The PDP can be expressed as:
 \begin{equation}   \label{eq2}
P_i[j]= P_i(jT_s),\;\;\; \forall j\in \left\{ 0,\cdots ,N_s -1 \right\}.
\end{equation}
where $N_s = \frac{T }{T_s}$ represents the total number of samples and $T$ denotes the maximum excess delay. The PDP samples are represented by binary values for each sampling time, i.e.,
 \begin{equation}   \label{eq3}
 x_{i,j} = I(P_i[j] \geq \eta),
\end{equation}
where $I: \mathbb{R} \rightarrow \mathbb{B} = \left\{ 0,1 \right\}$ is the indicator function and $\eta$ is a positive threshold level. By utilizing binary PDP values, implementation complexity and memory requirement of the TN can be decreased. Furthermore, utilizing a binary representation of the PDP can help handle high noise levels and improve localization performance \cite{gante2020deep}. Binary PDP values can be used to construct a PDP feature matrix $\textbf{X} = \left[ x_{i,j} \right] \in \mathbb{B}^{M \times N_s}$ which is transmitted to the BS for localization purposes.  

PDP data is collected in this manner for a predetermined grid of user terminal locations within the service area. The locations are indexed by $n\in \left\{ 1,\cdots ,{N} \right\}$, where ${N}$ is the total number of locations, and $\textbf{y}_{n} \in \mathbb{R}^{1 \times 2}$ represents the coordinates of the $n$th location in two-dimensional (2D) space. 
For each one of these locations, the corresponding PDP feature matrix $\textbf{X}_{n}\in \mathbb{R}^{M\times N_s}$ and position label $\textbf{y}_{n}\in \mathbb{R}^{1 \times 2}$ are collected as the input and output data for the TN. These matrices and vectors are collectively referred to as the BFF and are stored in the BS as a dataset $\mathcal{D} = \left\{ \textbf{X}_n,\textbf{y}_n \right\}_{n=1}^N $. This dataset is only used for training purposes. 

During the prediction phase, the same process is employed by the mobile users to collect temporal sequences of PDP feature matrices, say $\textbf{X}_{k} \in \mathbb{B}^{M \times N_s}$, where $k$ is now interpreted as a time index.
This collected data is then used as input to the trained TN model for trajectory estimation and path determination of the user terminal, as explained in the following sections.

Note that, the decision to use PDP data is driven by several factors related to the specific nature of our problem and the balance between computational efficiency and accuracy.

\section{Attention-Based Mechanism for BFF Trajectory Estimation}
In this section, we introduce a TN that leverages the BFF derived in the previous section to estimate users' trajectories. We first briefly discuss the modeling of user trajectories based on the BFFs, and then expose the desired TN architecture for trajectory estimation.
\subsection{Trajectory Modeling}

Before generating user trajectories or paths, we describe the method employed to generate a reliable dataset suitable for the application of BFF-based processing. The original dataset consists of pairs $\left\{ \textbf{X}_{k}, \textbf{y}_{k} \right\}$, where $\textbf{X}_{k} \in \mathbb{R}^{M\times N_s}$ is a matrix containing the received binary PDP samples and  $\textbf{y}_{k}$ is the vector of corresponding true positions. This data set is first converted into feature and label pairs, logging the ranges of these values for the purpose of verification and consistency. Additionally, the dataset is cleaned by removing unwanted time slots and positions without data. This systematic approach guarantees the integrity and usability of the dataset in subsequent analyses.

To generate user trajectories for tracking experiments, sequences of each trajectory elements are utilized based on specified data and path-related settings. The trajectories are initialized with true positions, grid dimensions, time steps, and various movement parameters for pedestrian and vehicular paths. We then process input labels to determine valid positions and generates dynamic trajectories. Dynamic paths for vehicles and pedestrians are iteratively constructed based on speed and direction adjustments. This approach ensures the consistency and reproducibility of paths by generating a unique identifier for each set of parameters. Paths are created for training, validation, and test sets, and can be stored and loaded efficiently. This structured approach facilitates robust tracking experiments through the systematic and reliable generation of paths.

The dataset can capture the various trajectories within the environment. Consequently, we represent each trajectory as $\textbf{Y}_{\eta} = [\textbf{y}^{(1)}, \textbf{y}^{(2)},\cdots ,\textbf{y}^{(L)}] \in \mathbb{R}^{1\times 2L}$, where $\eta$ is the trajectory index, $\textbf{y}^{(l)}$ represents an element of the trajectory associated with the feature matrix(i.e.,$\textbf{X}_{l}$), and $L$ is the path length. This detailed definition allows for a clear understanding of how trajectories are captured and represented within the dataset. The input of our models consists of different trajectories, represented as $\textbf{Y} = [ \textbf{Y}^{T}_{1},\textbf{Y}^{T}_{2},\cdots ,\textbf{Y}^{T}_{N_p}]^{\text{T}} \in \mathbb{R}^{ N_p \times 2L} $, where $N_p$ is the total number of trajectories \footnote{In practice, while the number of potential trajectories could be infinite due to the continuous nature of space, only a finite number of distinguishable paths can be identified due to the resolution limits of our approach. Moreover, physical environments have natural limitations and obstructions that reduce the number of feasible paths. Therefore to create a model that is computationally solvable within a reasonable time, we simplify the system by considering a finite set of representative paths.}. Each of these paths (i.e., $\textbf{Y}_{\eta}$), we consider a specific segment ranging from $l=1$ (the first element) to $l=\text{T}_{obs}$, and use it to predict the remaining segments from $l=\text{T}_{obs}+1$ to $L$. For each trajectory, $\left[ \textbf{y}^{(\text{T}_{obs}+1)} ,\cdots ,\textbf{y}^{(L)} \right]$ is considered as the target segment that should be estimated through the DL based models. We can express the concatenated matrix including the target segments from all trajectories as $\boldsymbol{\bar{\text{Y}}}$. The output of the DL based models, denoted as $\boldsymbol{\hat{\text{Y}}}$ provides an estimate of $\boldsymbol{\bar{\text{Y}}}$.
 
\subsection{Trajectory Estimation Using TNs}
\begin{figure}
\centering
\includegraphics[width=0.45\textwidth]{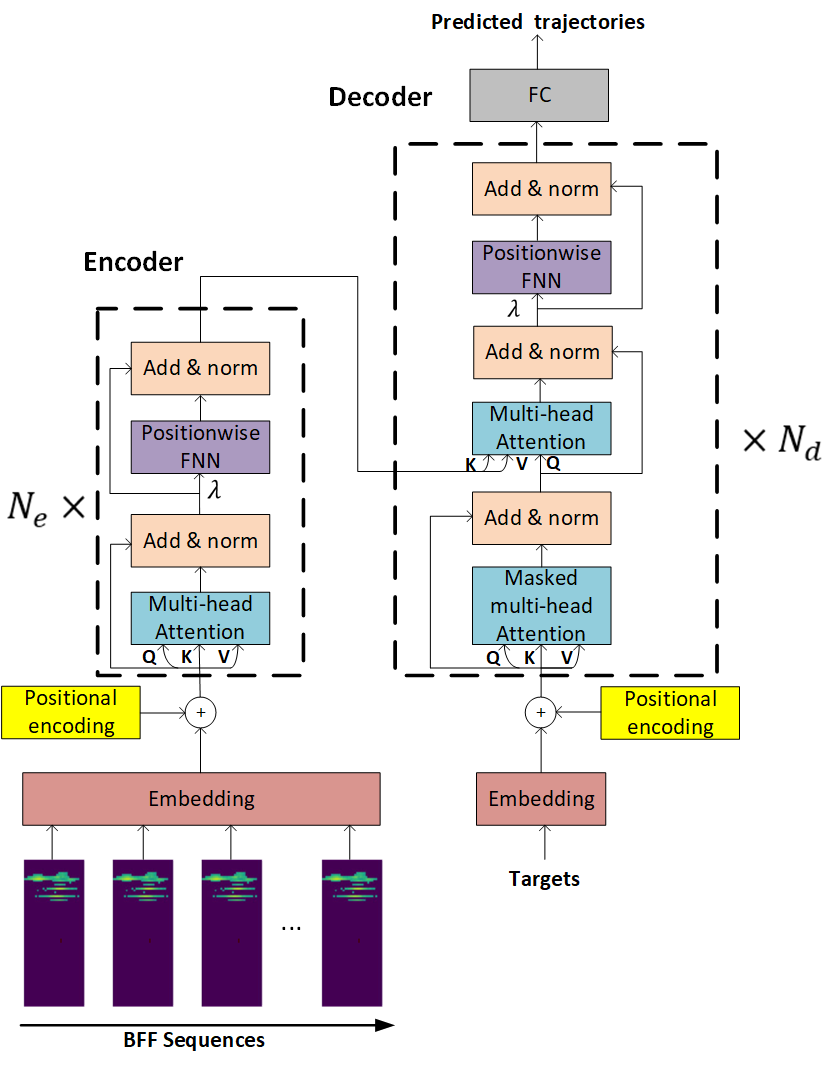}
\caption{\label{fig:frog} Representation of a TN model applied to BFF data.}
\label{transformer}
\end{figure}

The TN is a special type of neural network (NN) architecture that has found many applications in the learning-based processing of data with a sequential nature, such as in natural language processing (e.g., language translation, text-to-speech conversion, and more). While initially conceived for these types of applications, TN can also be applied to BFF sequences, which offer distinctive temporal signatures of the mmWave channel and represent the primary focus of this paper. The typical structure of the TN  consists of $N_{e}$ encoding layers and  $N_{d}$ decoding layers \cite{vaswani2017attention}, as illustrated in Fig. \ref{transformer}. The encoders process each element from the input sequence, which in this work consists of BFF measurements that captures the contextual surrounding, i.e., trajectory of a
user. The decoders then generate an output sequence which provides an estimate of the future trajectory of the user. Additionally, the TN incorporates the attention mechanism, which can focus to different parts of the input sequence at each step, based on their relevance to the output.

In our proposed TN implementation, the encoder block transforms the input sequence $\textbf{Y}$ into a latent state $\textbf{Z} = [ \textbf{Z}_{1}^{}, \textbf{Z}_{2}^{},\cdots ,\textbf{Z}_{N_p}^{} ]$; while given $\textbf{Z}^{}$, the decoder blocks generate the trajectory estimate $\boldsymbol{\hat{\text{Y}}}$. Note that, in an autoregressive manner, a previously generated trajectory element is used as an additional input to generate the subsequent element. In the TN, both the encoder and decoder blocks use stacked self-attention, position-wise, and fully connected (FC) layers. These layers can be seen in the left and right halves of Fig. \ref{transformer}, respectively.  For an in-depth discussion of the internal structures of the TN, we refer the reader to  \cite{he2016deep}.

The TN model mainly captures the dependencies in sequential data
and the non-linear attributes of BFFs through its attention module.
The latter relies on the parallel application of multiple self-attention
operations that allow the model to learn from subtle details of the input sequence. The attention mechanism is indeed a key feature of the TN that allows the network to focus on different parts of the input sequence when performing a task.  It computes a weighted sum of values, where the weight assigned to each value is obtained by means of a compatibility function of the query with the corresponding key, as explained below.

The self-attention operation takes as inputs the query matrix $\textbf{Q}$, the key matrix $\textbf{K}$, the value matrix $\textbf{V}$, and the dimension $d$ of the queries and keys. 
In this work, these matrices are obtained from the user's trajectory data. Matrix $\textbf{Q}$ can be derived from the current position's feature embedding. It represents the current position in the trajectory for which we want to predict the next position. Matrix $\textbf{K}$ can be derived from the historical or neighboring positions' feature embeddings. It serves to represent the reference position against which the current position (query) is compared. Matrix $\textbf{V}$ can be identical to $\textbf{K}$, or contain additional data associated with the keys, such as positional information.
The self-attention operation can be compactly expressed as  
\begin{equation}   \label{eq4}
\text{Attention} (\textbf{Q}, \textbf{K}, \textbf{V}) = \text{softmax}(\frac{\textbf{Q}\textbf{K}^{\text{T}}}{\sqrt{d}})\textbf{V}.
\end{equation}
It first computes the dot product of matrices $\textbf{Q}$ and $\textbf{K}$, which is scaled by a factor of $\frac{1}{\sqrt{d}}$. Subsequently, it applies a softmax function to the scaled dot product to generate normalized weights between 0 and 1. These weights are then applied to the input matrix ($\textbf{V}$). The scaling operation is particularly beneficial when $d$ is large, as it prevents the softmax function from entering regions where the network gradients are very small.

Rather than learning a single attention function, it has been found advantageous to map the queries, keys, and values multiple times in parallel to learn different information for each trajectory \cite{vaswani2017attention}. Known as multihead attention, this variant of the attention mechanism creates multiple sets of the query, key and value matrices for each input trajectory, allowing it to capture different aspects of attention.
In the multihead scheme, the attention function is applied to linearly transformed versions (also called projections) of the query, key and value matrices, represented by $\textbf{Q}\boldsymbol{\textbf{W}}_{i}^{Q}, \textbf{K}\boldsymbol{\textbf{W}}_{i}^{K}, \textbf{V}\boldsymbol{\textbf{W}}_{i}^{V}$ for $i = 1,\ldots ,h$, where the matrices $\boldsymbol{\textbf{W}}_{i}^{Q}$, $\boldsymbol{\textbf{W}}_{i}^{K}$, and $\boldsymbol{\textbf{W}}_{i}^{V}$ consist of learnable parameters and $h$ is the number of parallel attention-pooling modules. Following this, the outputs are concatenated and projected once more to determine the best and final value weights. These operations can be compactly expressed as follows,
\begin{subequations}
\begin{equation}   \label{eq5a}
\text{MultiHead}(\textbf{Q}, \textbf{K}, \textbf{V}) = \text{concat}(\text{head}_1,\cdots, \text{head}_h)\boldsymbol{\textbf{W}}^{O},\\
\end{equation}
\begin{equation}\label{eq5b}
\text{head}_{i} = \text{Attention}(\textbf{Q}\boldsymbol{\textbf{W}}_{i}^{Q}, \textbf{K}\boldsymbol{\textbf{W}}_{i}^{K}, \textbf{V}\boldsymbol{\textbf{W}}_{i}^{V}  ),
\end{equation}
\end{subequations}
where \emph{concat} refers to the concatenation operation (i.e., combining two or more vectors end-to-end to form a single longer vector) and the linear transformation $\boldsymbol{\textbf{W}}^{O}$ also consists of learnable parameters. By leveraging multi-head attention sub-layers as in (\ref{eq5a}), the TN is capable of simultaneously producing comprehensive latent features of the trajectory data.

In addition to the attention sub-layers, each layer within the above encoder and decoder architecture includes a fully connected feed-forward network. This network consists of two sequential linear transformations followed by a ReLU activation:
\begin{equation}
    \text{Feedforward}(\boldsymbol{\lambda}) = \text{ReLU}( \boldsymbol{\lambda}\boldsymbol{\textbf{W}}_{1}+\textbf{b}_{1})\boldsymbol{\textbf{W}}_{2} + \textbf{b}_{2},
\end{equation}
where $\boldsymbol{\lambda}$ represents the input vector at each position. The parameters $\textbf{W}_1, \textbf{W}_2, \textbf{b}_1$, and $\textbf{b}_2$ are specific to each layer but shared across all positions within that layer. This design choice ensures that while the linear transformations vary across layers, they remain consistent across positions within a layer. Moreover, we use learned embeddings to convert the input sequences and output sequences to vectors of certain dimension. We employ a fully connected (FC) block, which is designed to map the decoder's output directly to the probability distribution over the target position. The weights of the FC block are trained during the learning process to optimize the conversion from the decoder's output to the predicted next position.

Similar to other sequence-based models, this particular TN model uses learned embeddings to convert input and output tokens into vectors with a dimension of $\text{T}_{obs}$. A learned linear transformation function is also applied to the decoder output to generate predicted probabilities for the next token. The two embedding layers share a common weight matrix. In addition, the positional encoding technique \cite{vaswani2017attention} is used by the TN to help the model capture the sequential order of the input data. Unlike RNNs, the TN model does not have a natural way of encoding the order of elements in the sequence. Positional encoding provides a way to inject information about the sequence position into the input embeddings. Positional encoding works by adding a fixed vector to the embedding vector of each input token, thereby encoding the token's position within the sequence. The added vector is calculated based on a set of sinusoidal functions with different frequencies and phases \cite{gehring2017convolutional}. Specifically, the positional encoding for a token at position $pos$ and dimension $j$ is given by:
\begin{equation}   \label{eq7}
\text{PE}(pos, j) = 
\begin{cases}
  \sin({pos}/10000^{2j/\text{T}_{obs}}) & \text{if $j$ is even} \\
  \cos({pos}/10000^{2(j-1)/\text{T}_{obs}}) & \text{if $j$ is odd}
\end{cases}.
\end{equation}

As a final note, we emphasize that RNN-based methods such as LSTM process long data sequence element-by-element, causing long delays as signals traverse from the early to the late time steps. In contrast, the attention mechanism in the TN allows the entire input sequence to be processed in parallel, significantly reducing computational time. Furthermore, the TN does not suffer from the forgetting issues commonly associated with RNNs when dealing with long input sequences.
Unlike traditional recurrent networks, TNs can capture dependencies between data points without regard to their position in the sequence, making them more efficient for long-term trajectory prediction. In trajectory estimation, it is important to consider both short-term and long-term correlations in the target's movement \cite{gante2020deep}. The TNs' self-attention mechanism allows the model to capture relationships between distant time steps, which improves the accuracy of predicting future positions. The multi-head attention mechanism also enables the model to capture various patterns in the input data, which is useful for path determination. By attending to different parts of the trajectory simultaneously, the model can better understand the underlying structure of the target's path, leading to more accurate predictions.

\section{Path Determination Algorithm Principles}
In this section, we review the principles and properties of the RRT, RRT*, and iRRT* algorithms, which are sampling-based methods for path determination. We then explain how these algorithms can be effectively applied to the problem of path determination using the TN outputs produced by BFF data, emphasizing the novelty of our approach.

\subsection{Rapidly-exploring Random Trees}
The RRT algorithm is a method for determining feasible paths based on estimated trajectories (ETs). It works by randomly building a tree that explores the area created by ETs and connects the start and target positions. The RRT algorithm, summarized in Algorithm \ref{al1}, consists of three main steps: sampling, nearest neighbor, and steering. In each iteration, the algorithm samples a random position to find the nearest position in the existing tree, and tries to extend the tree towards the sampled positions. If the extension is successful, the new position is added to the tree. The algorithm terminates when the tree reaches the nearest area of the target position or a predefined number of iterations is reached \cite{lavalle1998rapidly}.

\begin{algorithm}
\caption{RRT Algorithm}
\begin{algorithmic}[1]
\Statex $\textbf{Require: } \text{ETs.}, \textbf{y}_{start}, \textbf{y}_{target}$
\State $\text{Tr}.\text{init}(\textbf{y}_{start})$;

\While{$\left\| \textbf{y}_{start} - \textbf{y}_{target} \right\|_2 \ge \epsilon$}

    \State $\textbf{y}_{rand} \gets \text{Sample}(\text{ETs.})$
    \State $\textbf{y}_{near} \gets \text{Near}(\textbf{y}_{rand}, \text{Tr})$
    \State $\textbf{y}_{new} \gets \text{Steer}(\textbf{y}_{rand}, \textbf{y}_{near}, \text{StepSize})$
    \If{$\text{CollisionFree}(\textbf{y}_{new})$}
        \State $\text{Tr}.\text{addPosition}(\textbf{y}_{near}, \textbf{y}_{new})$    
    \EndIf    
\EndWhile
\end{algorithmic}
\label{al1}
\end{algorithm}
The RRT* algorithm is an improved version of RRT that guarantees asymptotic optimality as the number of iterations increases. Besides implementing the same basic steps as RTT, RRT* also performs a rewiring step that updates the tree structure to reduce the path cost. The RRT* algorithm, summarized in Algorithm \ref{al2}, consists of four main steps: sampling, nearest neighbor, near neighbors, and rewiring. In each iteration, the algorithm samples a random position from the ETs, finds the nearest position in the existing tree, tries to extend the tree towards the sampled position, and checks the cost of the new position. If the cost is lower than that of existing positions in a neighborhood around the new position, the algorithm rewires the tree to connect the new position to the lower-cost positions \cite{elbanhawi2014sampling}.

\begin{algorithm}
\caption{RRT*}\label{alg:cap}
\begin{algorithmic}[1]
\Statex $\textbf{Require: } \text{ETs.}, \textbf{y}_{start}, \textbf{y}_{target}$
\State $\text{Tr}.\text{init}(\textbf{y}_{start})$;

\While{$\left\| \textbf{y}_{start} - \textbf{y}_{target} \right\|_2 \ge \epsilon$}

    \State $\textbf{y}_{rand} \gets \text{Sample}(\text{ETs.})$
    \State $\textbf{y}_{near} \gets \text{Near}(\textbf{y}_{rand}, \text{Tr})$
    \State $\textbf{y}_{new} \gets \text{Steer}(\textbf{y}_{rand}, \textbf{y}_{near}, \text{StepSize})$
    \If{$\text{CollisionFree}(\textbf{y}_{new})$}
        \State $\textbf{Y}_{near} \gets \text{NearC}(\text{Tr}, \textbf{y}_{new})$;
        \State $\textbf{y}_{min} \gets \text{ChooseParent}(\textbf{Y}_{near}, \textbf{y}_{near}, \textbf{y}_{new})$    
    \EndIf    
\EndWhile
\State $\text{Tr}.\text{addPosition}(\textbf{y}_{min}, \textbf{y}_{new})$
\State $\text{Tr}.\text{Rewire}$
\end{algorithmic}
\label{al2}
\end{algorithm}

\begin{figure}
\centering
\includegraphics[width=0.4\textwidth]{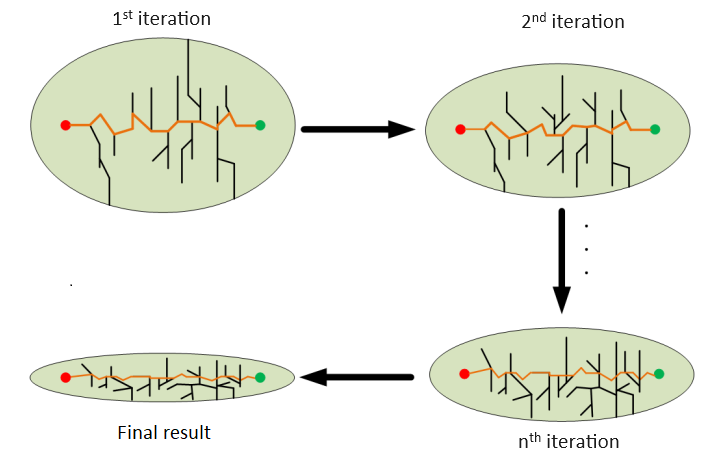}
\caption{\label{fig:frog}The iterative ellipsoidal shaping effect of iRRT* algorithm}
\label{fig5}
\end{figure}

The iRRT* algorithm is a variant of RRT* that improves the efficiency and optimality of path determination by focusing on an ellipsoidal area that contains the optimal path. iRRT* uses the current best path length and the distance between the start and target positions to define the shape and size of the ellipsoid. It then samples random states only from the ellipsoid, which reduces the search area and avoids unnecessary exploration. iRRT* also performs the same rewiring step as RRT* to update the tree structure and reduce the path cost. The iRRT* algorithm is summurized in Algorithm \ref{al3} \cite{gammell2014informed}. 
Before obtaining the initial path, iRRT* employs the same random sampling strategy as RRT*. Once the initial path has been found, the sampling area is adjusted to an elliptical shape with the starting position and target position as the foci. The length of the ellipse's major axis corresponds to the distance between the starting and target positions in the search tree. This adjustment makes the sampling more focused and improves the efficiency of the global path search as shown in Fig. \ref{fig5}.

\begin{algorithm}
\caption{iRRT*}\label{alg:cap}
\begin{algorithmic}[1]
\Statex $\textbf{Require: } \text{ETs.}, \textbf{y}_{start}, \textbf{y}_{target}$
\State $\text{Tr}.\text{init}(\textbf{y}_{start})$;
\State $c_{min}\gets\infty$; 
\State $\text{E}\gets \text{Ellipsoid} (\textbf{y}_{start}, \textbf{y}_{target}, \text{c}_{min})$; 

\While{$\left\| \textbf{y}_{start} - \textbf{y}_{target} \right\|_2 \ge \epsilon$}
\State $\textbf{y}_{rand} \gets \text{Sample}(\text{ETs})$; 
\State $\textbf{y}_{near} \gets \text{Near}(\textbf{y}_{rand}, \text{Tr})$;
\State $\textbf{y}_{new} \gets \text{Steer}(\textbf{y}_{rand}, \textbf{y}_{near}, \text{StepSize})$;
\If{$\text{CollisionFree}(\textbf{y}_{new})$}
    \State $\textbf{Y}_{near} \gets \text{NearC}(\text{Tr}, \textbf{y}_{new})$;
    \State $\textbf{y}_{min} \gets \text{ChooseParent}(\textbf{Y}_{near}, \textbf{y}_{near}, \textbf{y}_{new})$;
    \State $\text{Tr}.\text{AddPosition}(\textbf{y}_{min}, \textbf{y}_{new})$;
    \State $\text{Tr}.\text{Rewire}(\textbf{Y}_{near}, \textbf{y}_{new})$;
    \If{$\textbf{y}_{new} \in \text{GoalRegion}$}
        \State $\text{c}_{min} \gets \text{PathCost}(\textbf{y}_{new})$; 
        \State $\text{E} \gets \text{Ellipsoid}(\textbf{y}_{start}, \textbf{y}_{target}, \text{c}_{min})$; 
    \EndIf
\EndIf    
   
\EndWhile
\State $\text{Tr}.\text{AddPosition}(\textbf{y}_{min}, \textbf{y}_{new})$
\State $\text{Tr}.\text{Rewire}$
\end{algorithmic}
\label{al3}
\end{algorithm}

\subsection{Application of iRRT* to Path Determination
}

In more detail, the required operations in applying RRT, RRT*, and iRRT* to path determination using the ETs produced by BFF based TN model, can be expounded as follows:
\begin{itemize}
    \item \textbf{Sample}: Refers to randomly sampling positions in the ETs produced by the trained TN model.
    \item \textbf{Near}: Refers to selecting the sampling position on the search tree that is closest to $\textbf{y}_{rand}$ based on the Euclidean distance.
    \item \textbf{Steer}: Involves moving from the nearest position $\textbf{y}_{near}$ toward $\textbf{y}_{rand}$ by a fixed \textbf{StepSize} as expressed in (\ref{eq8}), resulting in a new position $\textbf{y}_{new}$.
    \item \textbf{CollisionFree}: Refers to verifying that the path from the parent position of $\textbf{y}_{near}$ to $\textbf{y}_{new}$ does not collide with any position outside the ETs.
    \item \textbf{NearC}: Refers to selecting all positions within a circle centered on $\textbf{y}_{new}$.
    \item \textbf{ChooseParent}: Involves selecting the parent position for $\textbf{y}_{new}$ among the positions in \textbf{NearC}.
    \item \textbf{AddPosition}: Refers to adding the newly generated position and its parent position to the search tree.
    \item \textbf{Rewire}: Involves updating the parent positions of all positions within the circle centred on $\textbf{y}_{new}$.
    
\end{itemize}
\begin{equation}\label{eq8}
\textbf{y}_{new} = \textbf{y}_{near} + \textbf{StepSize} \cdot \frac{\textbf{y}_{rand} - \textbf{y}_{near}}{\|\textbf{y}_{rand} - \textbf{y}_{near}\|}
\end{equation}

To generate the shortest possible path, we introduce a cost function that incorporates the Euclidean distance. Specifically, the total path cost is defined as, 
\begin{equation}\label{eq9}
    c_{total} = c (\textbf{y}_{new}) + \text{dist}(\textbf{y}_{new},\textbf{y}_{target}),
\end{equation}
where $c (\textbf{y}_{new})$ is the cost from the start position to the target position, and $\text{dist}(\textbf{y}_{new},\textbf{y}_{target})$ is the Euclidean distance from position $\textbf{y}_{new}$ to the target position $\textbf{y}_{target}$. The total path cost $c_{total}$ in (\ref{eq9}) is used to guide the iRRT* algorithm in selecting paths during the overall path determination process. It is computed as the sum of two critical components, balancing the already traversed path cost with the remaining distance to the target. This formulation ensures that the algorithm prioritizes paths that not only minimize the distance to the target, but also consider the quality of the path taken so far.

Unlike previous methods that use fixed or random sampling strategies, we use a data-driven approach to estimate the trajectories. Specifically, in our proposed approached, the trained TN model uses BFF sequences from the mmWave MIMO dataset to estimate user trajectories, which is followed by path determination using an RTT based scheme. We can apply anyone of RTT, RTT* or iRTT* to find the possible optimal user path from the start to the target position, while avoiding collisions with obstacles such as buildings, etc. Our approach extends the utility of BFF sequences by including spatial movement estimation. To process this data, we employ the TN with attention mechanisms, a first in this domain. The TN excels at modeling dependencies and capturing subtle patterns in data. Applying this architecture to trajectory estimation allows us to achieve high accuracy, surpassing the performance of conventional methods.
Once the trajectory is accurately estimated, we recommend using the iRRT* algorithm for precise path planning. iRRT* is particularly effective at finding optimal paths in large areas,, making it well-suited for our application. A key distinction in our approach is the integration of the iRRT* algorithm with the output of an attention-based TN. This integration enhances the performance of both trajectory estimation and path determination, providing more accurate and efficient results.

\section{Numerical Results}
In this section, we present the results of numerical experiments to evaluate the performance of various architectures for trajectory estimation and path determination. We use trained DL networks to learn the trajectories from BFF data, and then test them on new, unseen trajectories. We also use a sampling process that incorporates heuristic information to determine the optimal path to the target positions. The presentation is organized in three parts, respectively covering methodology, results for trajectory estimation, and results for path determination.

\subsection{Methodology}
Consistency is a crucial feature of any dataset intended to be used for training a ML model, as it enables the system to extract reliable information from its environment. To ensure consistency, the input data should be collected using an unchanging methodology. That is, both the transmission and reception procedures must remain constant in order to obtain accurate BFF data.

We utilize the Wireless InSite ray-tracing simulator to generate mmWave data \cite{Remcom}. This allows us to analyze the propagation of wireless signals in complex urban environments and assess the accuracy of our model. We consider a detailed 3D map of the New York University (NYU) region \cite{nework}, which includes BFF data from 160801 distinct two-dimensional locations (401 x 401 positions). In accordance with the findings in \cite{azar201328}, we employ the propagation parameters and the ray-tracing simulations, which are found to be consistent with experimental measurements. The propagation specifications outlined in Table \ref{ray} are derived from the experimental measurements detailed in \cite{gante2018data, gante2020deep}.
\begin{table}[]
\caption{Parameters for Ray-Tracing Simulations.
\label{long}}
\centering
\begin{tabular}{ |c|c|c| }
\cline{1-2}
\textbf{Parameters}              & \textbf{Values} \\
\cline{1-2}
Transmit Power         &   30 dBm    \\
\cline{1-2}
Carrier Frequency              &   28 GHz    \\
\cline{1-2}
Codebook Size       &   32 \\
\cline{1-2}
Tx. Antenna Gain &    24.5 dBi (horn antenna)\\
\cline{1-2}
Rx. Gain &    10 dBi \\
\cline{1-2}
Sampling Frequency &    20 MHz \\
\cline{1-2}
Noise Level &    6 dB \\
\cline{1-2}
Transmitter location &    (200 m, 200 m) \\
\cline{1-2}
\end{tabular}
\label{ray}
\end{table}
We add noise into the ray-tracing data using a log-normal distribution with a normal noise level of $6$ dB, unless otherwise indicated. Note that, the employed ray-tracing software lacks comprehensive support for MIMO antenna system with beamforming capabilities, a physically rotating horn antenna is utilized as the transmitter instead. The specifications for this antenna are detailed in \cite{gante2018data}.

\begin{figure}
\centering
\includegraphics[width=0.46\textwidth]{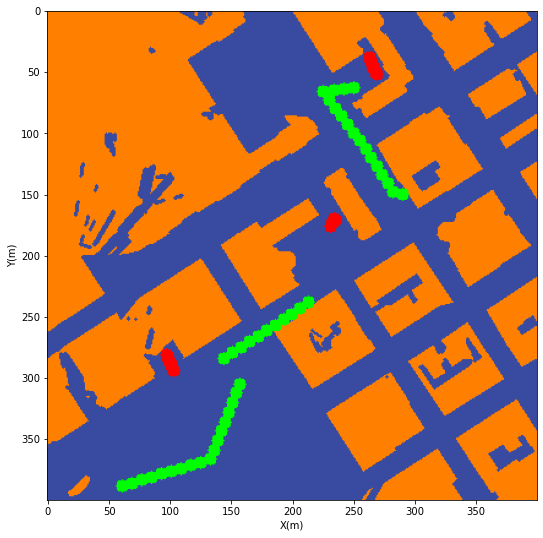}
\caption{\label{fig:frog}Examples of generated user trajectories in the NYU area. Areas valid for users, including vehicles and pedestrians, are indicated in blue. Invalid areas, such as buildings, are marked in orange. The movements of vehicles are shown in green, while those of pedestrians are shown in red}.
\label{fig2}
\end{figure}

The BFF data is obtained from the ray-tracing simulator by following the procedure detailed in Section II. To create BFF sequences for the training of the TN model and the evaluation of tracking based methods, i.e. trajectory estimation and path determination, two different types of synthetic user trajectories are produced, i.e.: vehicle-like and pedestrian-like sequences. The pedestrian-like sequences are designed to allow sudden stops or quick changes in direction at a low average speed of 5 km/h, while the vehicle-like sequences have a higher average speed of 30 km/h. The sequences in the dataset are sampled at a rate of 1 Hz, producing paths as exemplified in Fig. \ref{fig2}.
In more details, the paths are generated by simulating the motion of objects (e.g., pedestrians, vehicles) over time within a predefined environment. This involves defining the possible positions using a grid and creating both static and dynamic paths. Static paths represent stationary objects, while dynamic paths simulate movement based on probabilistic models. The movement is governed by parameters such as average speed, acceleration, direction changes, and probabilities of different motion types (e.g., stopping, changing direction). These paths capture short-term correlations by modeling immediate changes in position due to variations in speed and direction, reflecting local motion dynamics. Long-term correlations are embedded by considering the cumulative trajectory over extended time steps, capturing the object's overall behavioral patterns. The proposed architecture leverages these generated paths to learn these correlations, enabling accurate modeling of both rapid, short-term transitions and slower, long-term trends in the target's motion. This dual-focus approach ensures robust performance in tasks such as tracking and prediction, even though the underlying mechanisms may initially seem complex.

With regard to user trajectories, the training, validating, and testing data are selected from distinct subsets of trajectories in order to avoid memorization. In this paper, we utilize a comprehensive dataset comprising 432550 different trajectories in total. To ensure robust model evaluation and mitigate overfitting, the dataset is divided into three subsets: training, validation, and testing with the sizes of 75\%, 15\% and 20\% of the complete dataset, respectively. The training set is used to fit the model, allowing it to learn from the data. The validation set is employed to tune hyperparameters and make decisions on model selection. Finally, the test set provides an unbiased evaluation of the final model’s performance. This division ensures a balanced and thorough assessment of our model, contributing to the reliability and validity of our findings.
 
In our study, we have used a ray-tracing dataset, which is explicitly designed for a user scenario without interference. Therefore, our work assumes that the input data is associated exclusively with a pedestrian or a vehicle, and there is no interference from other targets.

\subsection{Results for  trajectory estimation}
Six different DL architectures from the literature that are appropriate for addressing the trajectory estimation problem have been implemented for comparison, namely: LSTM \cite{bai2018rfedrnn}, and temporal convolutional network (TCN) \cite{gante2020deep}, RNN, Conv1D \footnote{ To the best of our knowledge, this is the first instance where the RNN and Conv1D architectures have been employed for this specific type of trajectory data.}, GRU \cite{goodfellow2016deep}, and TN \cite{vaswani2017attention}. 
The choice of hyperparameters for these architectures was based on various factors, including: nominal values specified in the literature; baseline parameter values taken from reference models; parameter tuning/optimization using validation data;
model consistency (i.e., performance stability, reproducibility and robustness) as well as processing complexity. These considerations along with related experimental work\footnote{The tuning and final selection of hyperparameters was done through a grid search to ensure optimal performance and generalization. In particular, we optimized the learning rate to ensure smooth and stable convergence, and The tuning and final selection of hyperparameters was done through a grid search to ensure optimal performance and generalization. In particular, we optimized the learning rate to ensure smooth and stable convergence and tuned the dropout rate to balance overfitting with model expressiveness. The number of layers and attention heads were carefully chosen to achieve an optimal
trade-off between computational efficiency and model capacity. Additionally, the batch size was selected based on memory constraints and the need to balance gradient stability with generalization, while regularization parameters were adjusted to enhance generalization and prevent overfitting} ultimately led to the choice of hyperparameters outlined in Table \ref{t1}, which is key in determining their complexity.
The convergence of the models is ensured through multiple strategies. The Mean Squared Error (MSE) loss, which is used as validation metric for the model, is minimized during training, with convergence indicated by stabilization of the loss. Early stopping is employed to avoid overfitting, halting training if the validation loss does not improve for a set number of epochs. A learning rate scheduler adjusts the learning rate when loss stagnates, ensuring smooth convergence. Empirical results demonstrate consistent convergence up to 100 epochs, depending on the dataset's complexity.
In this regard, Table \ref{t2}\footnote{The results reported in Table \ref{t2} and other figures are obtained by taking the mean of multiple trials, using a Core(TM) i7-1065G7 CPU processor with 20 GB DDR4 RAM operating at 1.5GHz via Google's TensorFlow framework \cite{abadi2016tensorflow}. 
Furthermore, based on the results in Table \ref{t2}, we observed that the reported latency is accurate under the conditions tested.} provides a comparison of three significant attributes: model size in terms of learnable parameters, training time, and prediction time \cite{vaswani2017attention}.
As can be seen, TN models are faster than LSTM, TCN, and GRU since all of the inputs are taken into account at once. LSTM networks are more challenging to train than TN due to their considerably larger number of parameters. While ConvID and RNN yield faster training and processing time, their performance in trajectory estimation is significantly worst than the other methods, among which the TN offers the best performance, as we show below.

\begin{table}[]
\caption{Hyperparameters of the DL architectures used for trajectory estimation.\label{long}}
\centering
\begin{tabular}{ |c|c|c| }
\cline{1-2}
\textbf{Parameter}              & \textbf{Value} \\
\cline{1-2}
Number of units         &   (256, 64)    \\
\cline{1-2}
MLP layers              &   2    \\
\cline{1-2}
Regression output       &   2 neurons (2D position)\\
\cline{1-2}
Num. of training sequences &    320408\\
\cline{1-2}
Epochs                  &   Up to 100 \cite{caruana2000overfitting}\\
\cline{1-2}
Batch size              &   64 \\
\cline{1-2}
Optimizer               &   ADAM \cite{kingma2014adam} \\
\cline{1-2}
Learning rate           &   $10^{-3}$   \\
\cline{1-2}
Dropout                 &      0.01  \\
\cline{1-2}
Parallel attention layers                 &      $h = 2$  \\
\cline{1-2}
Num. of encoding/decoding layers       &     $N_{e}= N_{d} = 2$  \\
\cline{1-2}
\end{tabular}
\label{t1}
\end{table}

\begin{table}[]
\caption{Number of learnable parameters, training time, and prediction time of DL models, with sequence length 15 for the vehicle users. \label{long}}
\centering
\begin{tabular}{ |c|c|c|c| }
\cline{1-4}
\textbf{DL model}     & \textbf{L.parameters} & \textbf{T.time (hours)} & \textbf{P.Time (mins)} \\
\cline{1-4}
Conv1D              & $2.3\times10 ^{4}$           & $00:15:55$                   & $00:16$                     \\
\cline{1-4}
RNN          & $8.7\times10 ^{4}$           & $00:53:36$                   & $00:59$                      \\
\cline{1-4}
LSTM                & $35\times10 ^{4}$           & $3:41:29$                   & $07:03$                     \\
\cline{1-4}
TCN                & $37.5\times10 ^{4}$           & $3:19:40$                   & $06:35$ \\
\cline{1-4}
GRU                 & $26\times10 ^{4}$           & $1:44:11$                   & $02:11$                     \\
\cline{1-4}
TN & $9.2\times10 ^{4}$           & $1:10:50$                   & $01:27$ \\
\cline{1-4}
\end{tabular}
\label{t2}
\end{table}

\begin{figure}
\centering
\includegraphics[width=0.48\textwidth]{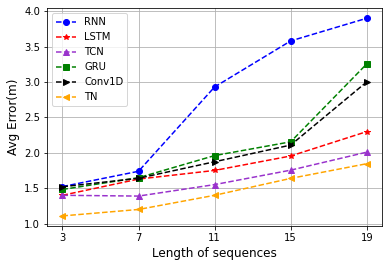}
\caption{\label{fig:frog} Comparison of trajectory estimation error of various models for vehicle trajectories}
\label{fig6}
\end{figure}

\begin{figure}
\centering
\includegraphics[width=0.48\textwidth]{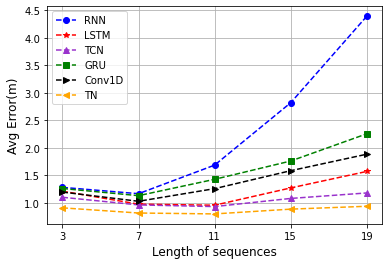}
\caption{\label{fig:frog}Comparison of trajectory estimation error of various models for pedestrian trajectories}
\label{fig7}
\end{figure}

Fig. \ref{fig6} and Fig. \ref{fig7} illustrate the accuracy achieved by various DL models for the trajectory estimation task in the case of car and pedestrian users, respectively.
The average error measures the distance between the predicted trajectory and the true trajectory, i.e.:
\begin{equation}
    \text{RMSE} = \sqrt{\frac{1}{N} \sum_{n=1}^{N}\left\| \boldsymbol{\bar{\text{y}}}_{n}-\boldsymbol{\hat{\text{y}}}_{n} \right\|^{2}},
\end{equation}
where $N$ is the total number of trajectories in the environment. The vector $\hat{\textbf{y}}_{n}$ corresponds to the predicted positions along a given trajectory, while $\bar{\textbf{y}}_{n}$ denotes the true positions for the same trajectory. In Fig. \ref{fig6} and Fig. \ref{fig7}, the average RMSE is plotted as a function of the sequence (i.e. trajectory) length. RMSE is a widely accepted metric to quantify the overall trajectory estimation error. While we acknowledge its limitations in capturing specific characteristics such as velocity and acceleration, our approach focuses on the path determination of each user. Since these characteristics are inherently reflected in the final path, RMSE provides a comprehensive evaluation of the model’s performance. The TNs clearly outperform other approaches and maintain high precision, even for long trajectories, by effectively leveraging the additional temporal information and capturing key features across the entire sequence. It is noteworthy that, TNs are able to achieve high levels of accuracy in trajectory estimation, on the order of a meter or less for pedestrians and about two meters for cars. We note that a significant proportion of the trajectories exhibit minor directional shifts, i.e., small changes in the direction of motion that occur frequently throughout the trajectory. Hence, the results illustrate the TN model’s superior performance in detecting these changes compared to other models. This could be explained by the TN reliance on the attention mechanism, which can focus and extract information from subtle details in sequential data, as discussed in Section III. In addition, TNs adapt to variations in moving speed by indirectly capturing velocity-related information through spatial and temporal features in the input data. The trained model can be applied to both slower, stable movements of pedestrians and faster, complex variations of vehicles, ensuring robust trajectory predictions across diverse motion dynamics without explicitly estimating velocity. In effect, the TN architecture is designed to process temporal dependencies at different scales, ensuring robust predictions
across a range of motion dynamics.

\begin{figure}
\centering
\includegraphics[width=0.48\textwidth]{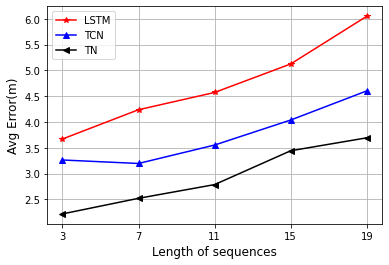}
\caption{\label{fig:frog}Comparison of 95th percentile of estimation error of different models for vehicle trajectories.}
\label{95vehicle}
\end{figure}

\begin{figure}
\centering
\includegraphics[width=0.48\textwidth]{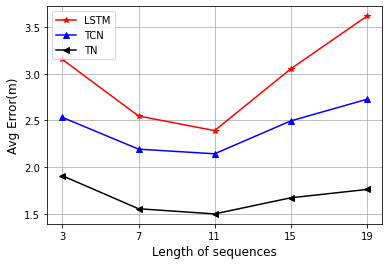}
\caption{\label{fig:frog}Comparison of 95th percentile of estimation error of different models for pedestrian trajectories.}
\label{95ped}
\end{figure}

To gain further insight into the comparative trajectory estimation accuracy of the three best DL models, i.e., TN, TCN and LSTM, Fig. \ref{95vehicle} and \ref{95ped} show plots of the $95^{th}$ percentile of their average RMSE versus sequence length for vehicle and pedestrian users, respectively, under a noise level of 6 dB. As can be seen, the TN model significantly outperforms the other two state-of-the-art methods in terms of this \footnote{According to this metric, the measured RMS remains below ($\le$) the plotted value for 95 percent of the simulated trajectories, while exceeding this value for the remaining 5 percent of trajectories. This metric provides a useful measure of accuracy across the various approaches, by ensuring that extreme values do not dominate the performance assessment while retaining most of the meaningful information.}. In Table \ref{t9}, we compare the performance of these three models for a sequence length of 7 and noise level $\sigma=6$ and $9 $ dB.
The 6 dB level simulates moderately noisy conditions typically encountered in urban environments, while the 9 dB level represents higher noise levels that might be found in more challenging environments such as industrial zones. This selection ensures that our method is rigorously tested under varying conditions to validate its robustness and generalization capability across different real-world scenarios.
The results point to the superior performance of the TN model under these conditions. Despite the increased noise level, the TN model consistently outperforms both the LSTM and TCN models. By incorporating noisy data during training, the TNs are able to generalize effectively and maintain consistent performance, as demonstrated by the experiments detailed in Table \ref{t9}\footnote{The parameter noise level represents the standard deviation of Gaussian noise applied to the values in dB, simulating a log-normal distribution. The dB scale here is relative to the logarithmically transformed power values of the signal.}. For handling
missing data, the TNs utilize self-attention to infer absent information, further supported by data imputation techniques, ensuring reliable trajectory estimation even in the presence of significant data gaps.

\begin{table}[h]
\caption{Comparing the trajectory estimation performance of the TN, TCN, and LSTM models for a sequence length of 7 at different noise levels.}
\centering
\begin{tabular}{ |c|c|c|c| }
\hline
\textbf{DL architecture} & \textbf{Noise level} & \textbf{Veh. AvgError(m)} & \textbf{Ped. AvgError(m)} \\
\hline
LSTM          & $\sigma=6$ dB & $1.63$  & $0.97$  \\
          & $\sigma=9$ dB & $2.2$  & $1.0$  \\
\hline
TCN           & $\sigma=6$ dB & $1.39$  & $0.96$  \\
           & $\sigma=9$ dB & $1.69$ & $0.98$\\
\hline
TN   &$\sigma=6$ dB & $1.2$  & $0.81$  \\
   & $\sigma=9$ dB & $1.41$ & $0.89$ \\
\hline
\end{tabular}
\label{t9}
\end{table}

\begin{figure}
\centering
\includegraphics[width=0.48\textwidth]{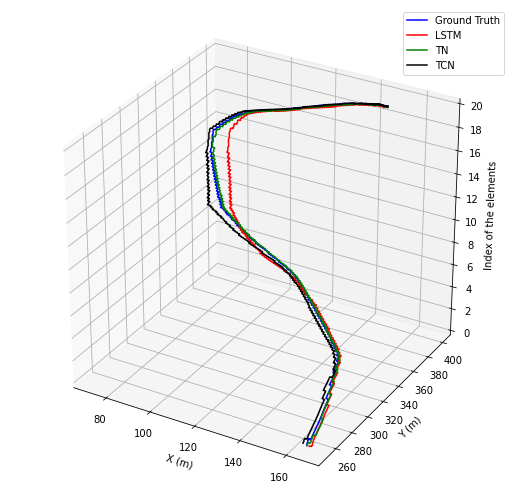}
\caption{\label{fig:frog}Comparison of trajectories predicted by LSTM, TCN and TN models, with ground truth trajectory exhibiting hybrid features.}
\label{ground_truth}
\end{figure}

In Fig. \ref{ground_truth}, we compare the prediction accuracy of three different DL architectures, i.e. proposed TN-based model, LSTM and TCN. In this example, the ground truth trajectory (referred to as hybrid) differs from those used previously in that it corresponds to a moving user with kinematic features in between those of a pedestrian and vehicle.  To enhance the visual presentation of the trajectories, the time dimension (i.e., element position in the sequence) is added along the z-axis. As observed in the figure, the trajectory predicted by the TN model is closer to the true trajectory than those predicted by the LSTM and TCN model.

The application of TN in trajectory estimation is particularly advantageous due to their ability to simultaneously process high-dimensional BFF data and capture long-range dependencies. Unlike RNN-based models, which may struggle with gradient decay and limited context windows, TN maintain global awareness of the entire sequence, ensuring accurate modeling of spatio-temporal correlations critical for trajectory prediction. The self-attention mechanism enables the prioritization of relevant features, which is essential to distinguish subtle motion patterns in diverse and noisy environments, further validating their suitability over traditional sequential models \cite{Su2025, Islam2023, Giuliari2020}.

The computational complexity of the TNs primarily arises from its self-attention mechanism, which scales quadratically with the sequence length $L$ and linearly with the model dimensionality $d$, i.e., according to $O(L^2d)$ \cite{vaswani2017attention}. To address this, we implement windowing techniques and efficient attention mechanisms, reducing computational overhead while maintaining prediction accuracy. Although computational complexity increases with larger datasets and more complex environments, the proposed strategies, such as sequence length limitation, efficient attention mechanisms, informed sampling, and parallelization, ensure that our methods remain scalable and efficient even in the presence of these challenges.

Despite the increased memory usage and processing time, the proposed TN remains an attractive option for real-time applications such as autonomous driving. This efficiency can indeed be achieved through the use of parallel processing and advanced optimization strategies, which significantly mitigate the computational overhead. Such an approach aligns with the strict timing requirements of autonomous systems, where rapid and reliable decision-making is critical for safety and operational efficiency.

\subsection{Results for Path Determination}
In this subsection, we compare the performance of three path determination algorithms, i.e., RRT, RRT*, and iRRT*, as applied to estimated trajectories derived from the trained BFF-based TN models. The simulation setup assumes a starting position at $\textbf{y}_{0} =$ (215, 193) m, and a target position at $\textbf{y}_{f} =$ (177, 46) m, while the \emph{StepSize} parameter is set to 5. The primary objective is to evaluate the algorithms' efficiency in terms of convergence to the target, path cost, and computational run time. The true trajectory refers to the optimal path that minimizes the distance and cost in (\ref{eq9}) from the start to the target position.

\begin{figure}
\centering
\includegraphics[width=0.48\textwidth]{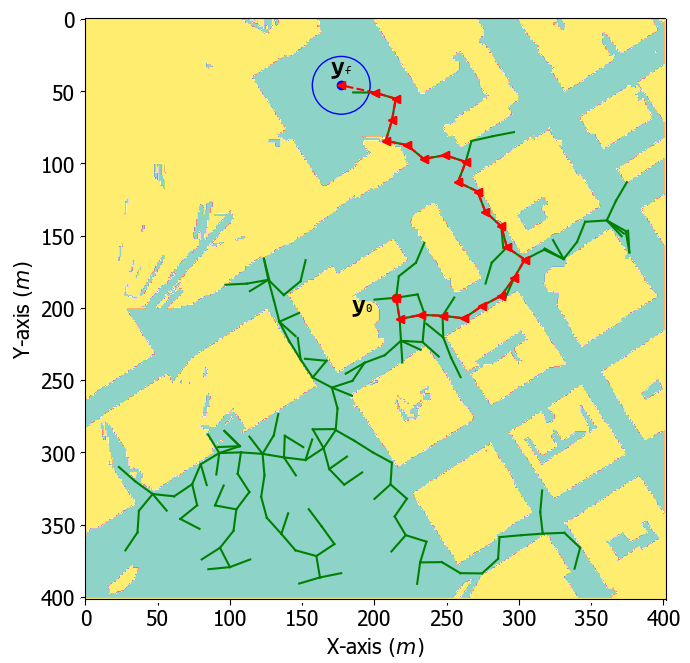}
\caption{\label{fig:frog} {Path discovery between two different positions using the RRT algorithm}.}
\label{fig8}
\end{figure}

\begin{figure}
\centering
\includegraphics[width=0.48\textwidth]{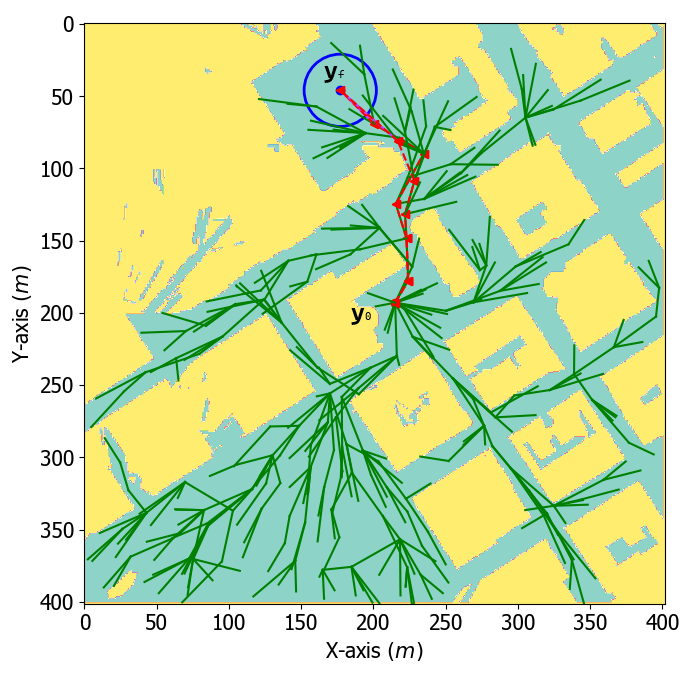}
\caption{\label{fig:frog} {Path discovery between two different positions in the network via the RRT* algorithm}.}
\label{fig9}
\end{figure}

\begin{figure}
\centering
\includegraphics[width=0.48\textwidth]{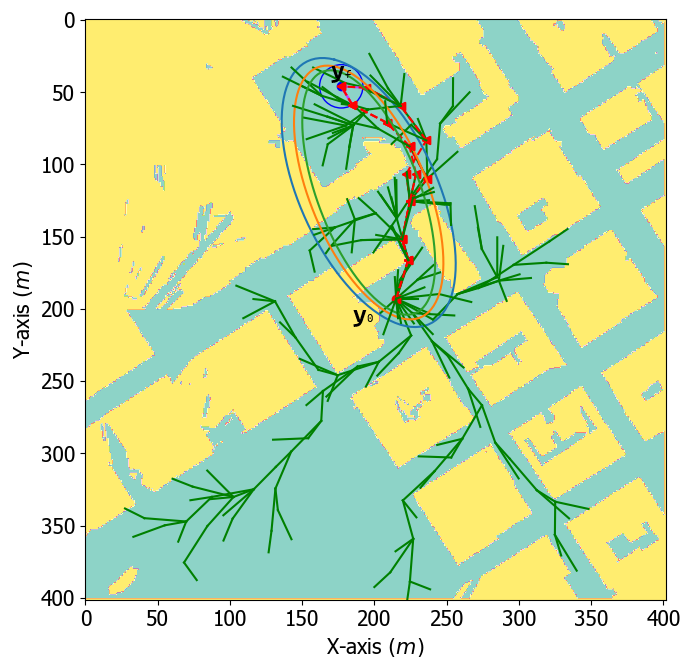}
\caption{\label{fig:frog}Path discovery between two different positions in the network using iRRT* algorithm}
\label{fig10}
\end{figure}

Fig. \ref{fig8}, \ref{fig9}, and \ref{fig10} illustrate the convergence of the RRT, RRT*, and iRRT* algorithms, respectively, as they progress from the start towards the target position. Each figure demonstrates the path generated by the corresponding algorithm during its run (green segments) along with the final optimized path (red dots). Notably, the iRRT* algorithm employs ellipsoidal refinement, which significantly enhances its convergence rate. This refinement method allows iRRT* to focus its search in a more directed manner towards the target, reducing unnecessary exploration, thus leading to faster convergence. Consequently, iRRT* demonstrates superior performance compared to RRT and RRT*, achieving lower path costs and reduced run times. The efficiency of iRRT* is evident as it consistently finds shorter paths with fewer iterations.
Specifically, the estimated trajectories from the TN are used as input to the iRRT* path determination algorithm. The key idea is that the predicted future positions of the users guide the path determination process. In the wireless environments, accurate trajectory estimations allows path determination algorithms to generate a more realistic and feasible paths, by anticipating the user's future movements and avoiding obstacles accordingly.
Thus, the integration of trajectory estimation into iRRT* enhances the adaptability and robustness of the overall system, particularly in scenarios where future positions must be considered to avoid dynamic obstacles or optimize paths.

Table \ref{t3} presents a comparison of the performance metrics of RRT, RRT*, and iRRT*. The table highlights the average path cost and run time, after 1000 iterations. Limiting the number of iterations  to 1000 ensures that the response time is acceptable for near real-time path determination.
The results indicate that iRRT* not only outperforms the other algorithms in terms of lower path cost but also shows a significantly lower run time. This improvement is attributed to the ellipsoidal refinement technique in iRRT*, which ensures a more precise and efficient path determination process. The result underscore the advantages of iRRT* in performing the final path determination from BFF data, making it a superior choice for applications requiring rapid and accurate path determination.
The proposed architecture and trained models can therefore provide invaluable support to autonomous drones, delivery robots, and automated systems in navigating dynamic and complex environments by enabling precise trajectory estimation and adaptive path detemination.

\begin{table}
\caption{Comparison of path cost and run time for RRT, RRT* and iRRT* algorithm after 1000 iterations.\label{long}}
\centering
\begin{tabular}{ |c|c|c|c| }
\cline{1-4}
    \textbf{Items}         & RRT        & RRT*    & iRRT*    \\
\cline{1-4}
Total Path Cost (m)     & 210          & 192       & 178        \\
\cline{1-4}
Run Time (sec) & 14.81          & 24.61       & 11.7        \\
\cline{1-4}
\end{tabular}
\label{t3}
\end{table}

For the iRRT* algorithm, the complexity is influenced by the search space size and the number iterations needed for convergence. Informed sampling significantly improves scalability by reducing the required number of iterations, enabling effective performance in complex radio environments with diverse multipath characteristics and higher user densities.  
The performance of the algorithm is closely tied to the accuracy of trajectory estimation by the TNs, as higher estimation accuracy enhances the algorithm's ability to identify optimal paths. While theoretically evaluating accuracy against the absolute optimal path requires exhaustive computation of all possible paths—an infeasible approach in most scenarios—we focus on balancing computational efficiency and path quality.

\section{Conclusion}
In this paper, we proposed and investigated a novel DL architecture suitable for trajectory estimation and path determination in a wireless network. In the first stage, a TN based on attention mechanisms is developed to predict the trajectories of users using BFF sequences captured in a mmWave MIMO outdoor system. In the second stage, a novel algorithm based on iRRT* is employed to determine the optimal path using estimated trajectories derived from BFFs in the first stage. A key novelty of our approach is the integration of trajectory estimates into the iRRT* algorithm, enabling dynamic and adaptive path determination. This approach leverages the trajectory estimation to improve the quality of the final path determination, addressing the limitations of static or pre-determined inputs. This enhancement significantly lowers computational overhead while preserving or improving path accuracy, as demonstrated by quantitative results. Indeed, simulation outcomes confirm that the proposed trajectory estimation approach outperforms existing methods in the literature and generalizes effectively to new, previously unseen trajectories. Additionally, the proposed iRRT* algorithm enables rapid and precise path determination.

The findings of this research can be applied to autonomous drones, delivery robots, and similar systems. While these applications are highly relevant, the current study primarily focuses on the fundamental model and its performance evaluation. Future work could expand on the practical implementation of these applications by specifying detailed conditions and real-world scenarios. Additionally, this paper mainly evaluates the overall performance of the TN model and does not extensively explore its internal dynamics, such as variations in time steps and feature dimensions \cite{Goodfellow2016, Rogers2021}. Investigating these aspects in future studies could provide deeper insights into the model’s behavior and potential optimizations.

\bibliographystyle{IEEEtran}
\bibliography{sample.bib}

\begin{IEEEbiography}[{\includegraphics[width=1in,height=1.25in,clip,keepaspectratio]{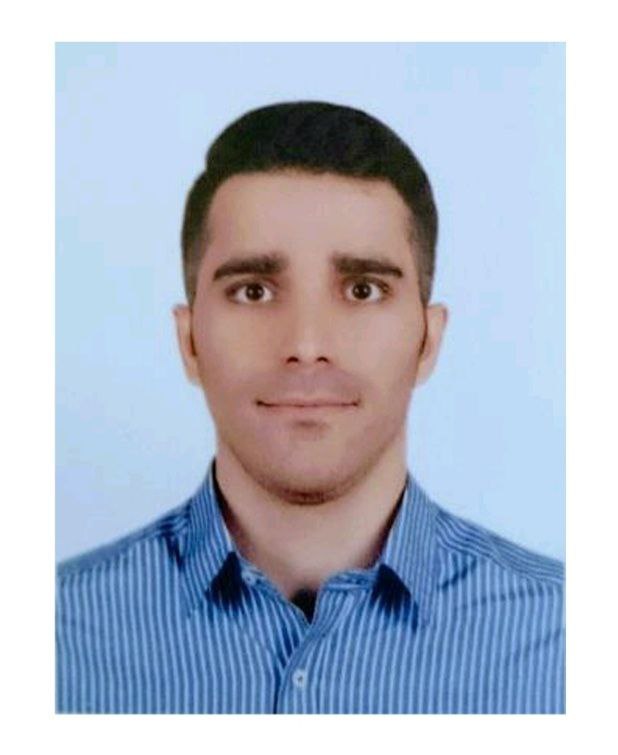}}]
{Mohammad Shamsesalehi}
earned his B.Eng. in Electrical Engineering from Jundi Shapour University of Technology, Dezful, Iran, in 2016, followed by an M.Sc. in Wireless Communications and Signal Processing from Qom University of Technology, Qom, Iran, in 2019. Presently, he is a Ph.D. candidate in Communication Systems at K.N. Toosi University of Technology, Tehran, Iran. His research is focused on leveraging machine learning approaches to address challenges in massive MIMO and millimeter-wave communication systems.
\end{IEEEbiography}
\begin{IEEEbiography}[{\includegraphics[width=1in,height=1.25in,clip,keepaspectratio]{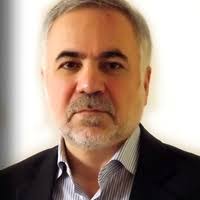}}]
{Mahmoud Ahmadian Attari}
is a Professor at the Department of Electrical   Engineering, K. N. Toosi University of Technology, Iran. He received his M.Sc. degree in Electrical Engineering from the University of Tehran, Iran, in 1977 and his Ph.D. degree in Digital Communication Systems from the University of Manchester in 1997. His research interests include coding theory and cryptography.
\end{IEEEbiography}
\begin{IEEEbiography}[{\includegraphics[width=1in,height=1.25in,clip,keepaspectratio]{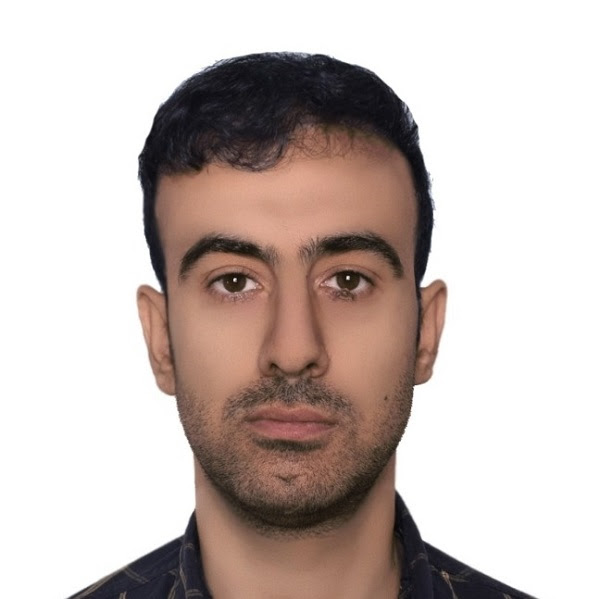}}]
{Mohammad Amin Maleki Sadr}
received his B.Sc. degree from the University of Isfahan, Iran, and his M.Sc. and Ph.D. degrees from K.N. Toosi University of Technology, Tehran, Iran, in 2012, 2014, and 2018, respectively, all in Electrical Engineering. He served as a Postdoctoral Research Fellow in the Department of Electrical and Computer Engineering at McGill University, Canada, during 2020-2021. Following this, he continued his postdoctoral research in the Department of Statistics and Actuarial Science at the University of Waterloo, Canada, from 2021 to 2023. His research interests are broad and include Deep Learning, Networking for Large Language Models, Anomaly Detection, Localization, Signal Processing, and innovations in Beyond 5G wireless communication technologies.
\end{IEEEbiography}
\begin{IEEEbiography}[{\includegraphics[width=1in,height=1.25in,clip,keepaspectratio]{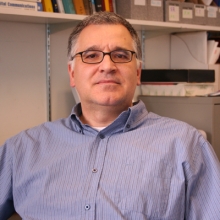}}]
{Benoit Champagne}
received the B.Ing. degree in Engineering Physics from the École Polytechnique de Montréal in 1983, the M.Sc. degree in Physics from the Université de Montréal in 1985, and the Ph.D. degree in Electrical Engineering from the University of Toronto in 1990. From 1990 to 1999, he was an Assistant and then Associate Professor at INRS-Telecommunications, Université du Quebec, Montréal. In1999, he joined McGill University, Montreal, where he is now a Full Professor in the Department of Electrical and Computer Engineering; he also served as Associate Chairman of Graduate Studies in the Department from 2004 to 2007. His research focuses on the study of advanced algorithms for the processing of communication signals by digital means. His interests span many areas of statistical signal processing, including detection and estimation, sensor array processing, adaptive filtering, and applications to broadband communications and audio processing, where he has co-authored nearly 250 referred publications. His research has been funded by the Natural Sciences and Engineering Research Council (NSERC) of Canada, the “Fonds de Recherche sur la Nature et les Technologies” from the Govt. of Quebec, as well as some major industrial sponsors, including Nortel Networks, Bell Canada, InterDigital, and Microsemi.
\end{IEEEbiography}

\end{document}